\documentclass[iop, apjl, twocolappendix, twocolumn]{aastex631}
\usepackage{graphicx}
\usepackage{bm}
\usepackage{amsmath,amssymb}
\usepackage{natbib}
\usepackage{float}
\usepackage{units}
\usepackage{color}
\usepackage[caption=false]{subfig}
\newcommand{\citeg}[1]{\citep[e.g.,][]{#1}}
\newcommand{\yr}{\,{\rm yr}}
\def\xiM{\xi_{\rm M}}
\newcommand{\Eq}[1]{Eq.~(\ref{#1})}
\usepackage{hyperref}
\def\Brms{B_{\rm rms}}
\newcommand{\m}{\,{\rm m}}
\newcommand{\s}{\,{\rm s}}

\begin{document}
\title{Confronting the neutron star population with inverse cascades}
\author{Nikhil Sarin}
\affil{Nordita, Stockholm University and KTH Royal Institute of Technology \\
Hannes Alfvéns väg 12, SE-106 91 Stockholm, Sweden}
\affil{Oskar Klein Centre for Cosmoparticle Physics, Department of Physics,
Stockholm University, AlbaNova, Stockholm SE-106 91, Sweden}
\email{nikhil.sarin@su.se}

\author{Axel Brandenburg}
\affil{Nordita, Stockholm University and KTH Royal Institute of Technology \\
Hannes Alfvéns väg 12, SE-106 91 Stockholm, Sweden}
\affil{The Oskar Klein Centre, Department of Astronomy, Stockholm University, AlbaNova, Stockholm SE-106 91, Sweden}

\author{Brynmor Haskell}
\affil{Nicolaus Copernicus Astronomical Centre, Polish Academy of Sciences, Bartycka 18, 00-716 Warsaw, Poland}

\shortauthors{Sarin et al.}

\begin{abstract}
The origin and evolution of magnetic fields of neutron stars from birth has long been a source of debate. 
Here, motivated by recent simulations of the Hall cascade with magnetic helicity, we invoke a model where the large-scale magnetic field of neutron stars grows as a product of small-scale turbulence through an inverse cascade. We apply this model to a simulated population of neutron stars at birth and show how this model can account for the evolution of such objects across the $P\dot{P}$ diagram, explaining both pulsar and magnetar observations.
Under the assumption that small-scale turbulence is responsible for large-scale magnetic fields, we place a lower limit on the spherical harmonic degree of the energy-carrying magnetic eddies of $\approx 40$. Our results favor the presence of a highly resistive pasta layer at the base of the neutron star crust.
We further discuss the implications of this paradigm on direct observables, such as the nominal age and braking index of pulsars.
\end{abstract}

\section{\label{sec:intro}Introduction}
Neutron stars harbor the strongest known magnetic fields in the universe, with the large-scale poloidal field strength $B_{\rm p}$ in the so-called magnetars exceeding values of $B_{\rm p}\approx 10^{15}{\rm G}$.
The magnetic field strength plays a crucial role in determining the observational properties of a neutron star and whether, for example, it will be seen as a standard radio pulsar or a magnetar~\citeg{borghese_2020}.
Furthermore, the large-scale dipolar field is thought to be primarily responsible for the spindown of neutron stars~\citep{Ostriker1969}, leading to a spindown law of the form $\dot{\Omega}\approx \Omega^n$, where $\Omega=2\pi/P$ is the spin frequency of the star with $P$ being the rotation period, and $n$ is the so-called braking index that can be observationally constrained from the combination $n=\ddot{\Omega}\Omega/\dot{\Omega}^2$. 
For dipole magnetic spindown, one has $n=3$, with deviations from this value expected if other mechanisms are driving the spindown (e.g., $n=1$ for winds, and $n=5$ for gravitational wave emission from quadrupolar `mountains').

In this picture, neutron stars are assumed to be born rapidly rotating, in the upper left corner of the $P\dot{P}$ diagram, and evolve towards the bottom right corner and the so-called `death' line as they spin down with a constant magnetic field. 
Observed braking indices are, however, often different from $n=3$, and there is significant observational and theoretical evidence for magnetic field evolution during the lifetime of a neutron star; see~\cite{Igoshev+21} for a recent review. 
Therefore, understanding neutron star formation and evolution mechanisms is essential to understanding the observed neutron star population.

The origin of neutron star magnetic fields is, however, still debated. 
It is generally agreed that the fossil field inherited from the progenitor star is insufficient to explain the strongest neutron star fields; therefore, some field amplification is necessary. 
Most models invoke a large-scale dynamo at birth, powered by convection or differential rotation~\citep{Thompson+Duncan93}.
The source of the turbulence driving the dynamo comes from neutrino-driven convection.
An additional source of turbulence is the magnetorotational instability
\citep{Guilet22, RS22}, which draws energy from the differential rotation, i.e., ultimately from potential energy.
This instability requires the presence of a magnetic field, which is accomplished through a positive feedback loop whereby magnetic energy can be amplified further by dynamo action \citep{BNST95}.
Other possible dynamo mechanisms tap energy through fallback accretion \citep{Barrere22} or through precession \citep{Lander21}.
In all those cases, small-scale and large-scale dynamo action appear together. 
In this Letter, we define a large-scale field as the part contained in the dipole component, while the small-scale field is the remainder.

Magnetic field amplification can also occur later in the life of a neutron star due to the re-emergence of a buried magnetic field due to Ohmic dissipation \citep{Muslimov96, Ho11} or Hall drift \citep{GC15}. 
Most of these models of magnetic field evolution in neutron stars assume initially low-order multipoles~\citep{GH18, Gourgouliatos+20}, although the true magnetic field structure is a matter of considerable debate~\citep{Igoshev+21}. 
Alternatively, the magnetic field could be predominantly of small scales at a very
young age of the neutron star, but the field could then undergo what is
known as inverse cascading \citep{WH09, WH10, Igoshev+21b}.
The spectral magnetic energy at small wavenumbers or low multiples
would then increase with time rather than decrease.
This idea was advanced in a recent paper \citep[][hereafter B20]{Bra20},
where it was shown that the resulting large-scale field, $B_{\rm LS}$,
increases approximately linearly with time, typically by three orders of magnitude in the models of B20, while the thermal emission continues to decrease.

\begin{figure*}[ht!]
\centering
  \begin{tabular}{cc}     
        \includegraphics[width=0.92\textwidth]{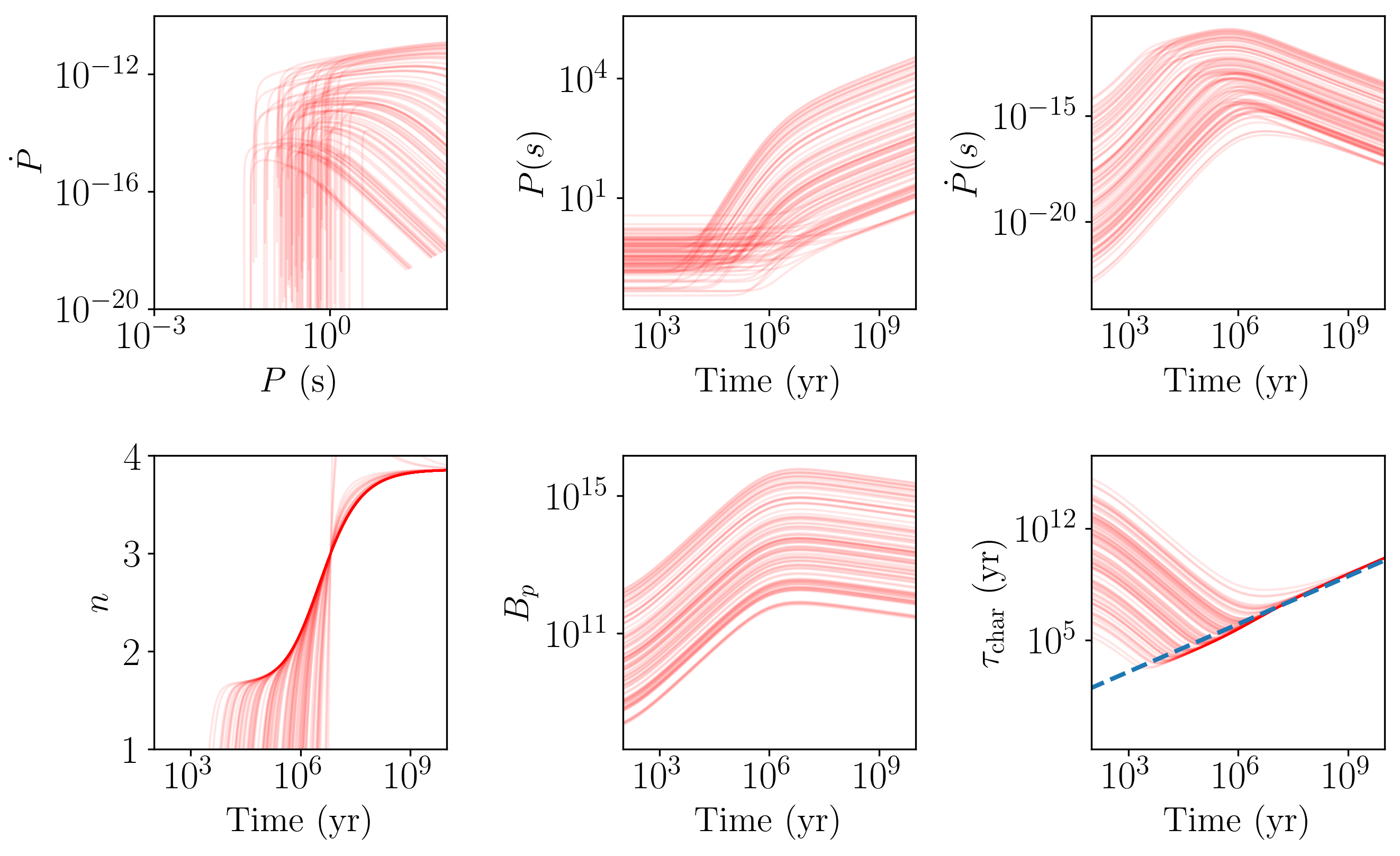} \hspace{-.4cm}
  \end{tabular}
  \caption{Evolution of neutron star properties for varying initial conditions (red curves). From left to right, the top panels show the evolution across the $P\dot{P}$ diagram, and then the evolution with time of the period, and the period derivative. The bottom panels show the evolution with time of the observed braking index, the large-scale magnetic field, and the characteristic age.} 
\label{fig:timeevolution}  
\end{figure*}

In this Letter, we investigate the implications of the scenario proposed in B20
on a simulated population of neutron stars and observables such as the braking index, characteristic age $\tau_{\rm char}$, and the viability of this idea to explain the neutron star population.
We begin in Section~\ref{sec:b_increase} discussing the motivation for a model where the magnetic field grows from a small-scale turbulent field at birth to a large-scale field.
In Section~\ref{sec:model}, we describe the spin evolution of a neutron star under this hypothesis and perform simulations showcasing the evolution of a population of neutron stars for various observables in Section~\ref{sec:simulation}.
We discuss implications of our results and conclude in Section~\ref{sec:conclusions}.
\section{Growth of large-scale field}\label{sec:b_increase}
A proto-neutron star is typically fully convective at birth and remains so for tens of thousands of turnover times~\citep{Epstein79}, corresponding to a time of $\mathcal{O}(\unit[1]{min})$~\citep{Burrows+Lattimer86}. 
It is commonly assumed that, within this time, the progenitor magnetic field is destroyed and then regenerated by a dynamo~\citep{Thompson+Duncan93}.
However, given the limited time between the collapse of the progenitor star and the end of the convective period, it is conceivable that only a small-scale magnetic field with scales comparable to the size of the turbulent eddies is generated. 
At the neutron star surface, the scale of such a field can be estimated by the density scale height $H$, which we expect to coincide roughly with the thickness of the crust of $\approx 5\%$ of the neutron star radius~\citep{Cumming+04}.
This corresponds to a typical wavenumber $k=1/H$, and thus to a typical spherical harmonics degree of $\ell=(H/R)^{-1}\approx 20$, where $R$ is the radius of the neutron star.
Note, however, that the thickness of the crust will depend on the star's mass, the details of the equation of state, and the scale height will be some fraction of this thickness. Therefore, we will explore a range $10 \leq \ell_{\rm max} \leq 40$. 

Eventually, when the neutron star solidifies, the magnetic field in the crust is still randomly distributed in space. It then decays in a way analogous to a turbulent decay, but here the dynamics are governed by the Hall cascade~\citep{GR92}. In this case, similar to the case of ordinary magnetohydrodynamic (MHD) turbulence, the magnetic field undergoes an inverse cascade, where not only the peak of the magnetic energy spectrum moves toward smaller $k$, but also the spectral energy at the smallest wavenumber {\em increases} in time.
Numerical simulations in B20 found the large-scale field $B_{\rm LS}$ to increase with decreasing rms magnetic field strength, $\Brms$, like
$B_{\rm LS}\propto\Brms^{-5}$; see his Figure~12(b).

In the Hall cascade, $\Brms$ decays with time like $\Brms\propto t^{-p/2}$, where $p=2/5$ is the decay exponent of the magnetic energy for a helical field (B20).
Thus, 
\begin{equation}\label{eq:BLS1}
B_{\rm LS}\propto\Brms^{-5}\propto t,
\end{equation}
so the large-scale magnetic field increases linearly with time; see~\cite{SM} for supplemental material showing that the growth of $B_{\rm LS}$ is actually closer to $t^{1.2}$, although this minor departure from the expected linear growth will be ignored here.
This linear growth applies to the case when the magnetic field is fully helical (see Runs~E and F in Figure~12 of B20).
However, it also applies to the case of nearly nonhelical fields (see Runs~C and D in Figure~12 of B20) because the magnetic helicity is nearly unchanged, but the magnetic energy decays, so the ratio increases, making the magnetic field in the end nearly fully helical; see \cite{Tevzadze+12} for similar behavior in the MHD case.

The increase of $B_{\rm LS}$ continues until the largest scale in the system has been reached.
Mathematically, the scale increases with $t$ like $\xiM\propto t^q$, where $q=2/5=0.4$ in the helical case, and $4/13\approx0.31$ in the nonhelical case \citep{Bra23}.
When $\xiM$ becomes comparable to $R$, we must adopt global spherical geometry.

So far, inverse cascading has not yet been seen in global simulations. 
There are multiple reasons for this.
First, it helps to initialize the simulations with a magnetic field having a broken power law energy spectrum with a proper inertial range
and a sufficiently steep subinertial range.
Second, large numerical resolution is required to obtain inverse cascading, at least in the nonhelical case (B20 used $1024^3$ mesh points).
The global simulations of \cite{Gourgouliatos+20}, for example, did not have power laws, extended only to spherical harmonic degrees of eighty, and did not include magnetic helicity, which is the case considered here. 
\cite{Dehman+23} also presented global simulations with initially complex magnetic field configurations, but again not with power-law initial fields or with helicity.

In the absence of suitable global simulations, we argue here that
we can substitute the spherical harmonics representation of the Laplacian,
$\ell(\ell+1)/R^2$, by just $k^2$, as in Cartesian geometry.
Simplifying this further to $k\approx\ell/R$, and using $k\approx\xiM^{-1}\propto t^{-q}$ with $q=2/5$ in the helical case, we have
\begin{equation}\label{eq:ell}
\ell\approx kR\propto R t^{-q}.
\end{equation}
As argued above, the initial value of $\ell$ is expected to be somewhere around $\ell_{\rm max}\approx20$, and will then decrease to $\ell=1$ like
\begin{equation}\label{eq:ell2}
\ell=\ell_{\rm max}\,(t/\tau_1)^{-q},
\end{equation}
where $\tau_1$ marks the start of the decay.
Thus, we have $\ell=1$ at $t\equiv\tau_2=\tau_1\ell_{\rm max}^{1/q}$.

After that time, the inverse cascade stops, and then even the large-scale field can only decay.
This decay is expected to be either exponential or in a power-law fashion.
We assume here the latter and thus propose
\begin{equation}\label{eq:bevolution}
B_{\rm LS}^2=B_{\rm p,0}^2\frac{(1 + t/\tau_1)^{2}}{(1 + t/\tau_2)^{m}}
\end{equation}
for the time evolution of the large-scale magnetic field,
where $B_{\rm p,0}$ is the initial large-scale poloidal (dipolar $\ell=1$) component of the neutron star magnetic field.
Here, $m$ dictates
the slope of the large-scale field decay, which is not well constrained.
Finally, let us note that the value of $\tau_1$ is linked, in the simulations of B20, to the diffusive timescale $t_{\rm d}\approx \xiM^2/\eta$, where $\eta$ is the magnetic resistivity.
Low values of $\tau_1$ generally correspond to a high resistivity, as we will discuss in the following sections.
\section{Spin evolution}\label{sec:model}
With the behavior of the magnetic field described above, we can
continue to describe the evolution of the neutron star spin.
For the sake of simplicity, we start with the standard vacuum spin-down
formula~\citep{Ostriker1969} ignoring other spin-down mechanisms such
as gravitational-wave emission.
Using Equation~\eqref{eq:bevolution}, our neutron star spin-evolution is then governed by
\begin{equation}\label{eq:spinevolution}
\dot{\Omega} = - \frac{B_{\rm p,0}^2 R^{6} \Omega(t)^{3}}{6c^3 I}\frac{(1 + t/\tau_1)^{2}}{(1 + t/\tau_2)^{m}}.
\end{equation}
Here, $c$ is the speed of light, and $I$ is the moment of inertia of the neutron star.
We emphasize that the above model is effectively the common vacuum-dipole spin-down equation~\citep{Ostriker1969} except for accounting of the evolution of the large-scale magnetic field following \Eq{eq:bevolution}. 
Under the assumption that the magnetic field evolves from small scales to large scales in a fully helical manner, the timescales $\tau_{1}$ and $\tau_{2}$ can be related to each other via
\begin{equation}
\tau_{2} = \tau_1 \ell_{\rm max}^{5/2}.
\end{equation}
Here, $\ell_{\rm max}$ can be estimated as the inverse ratio of the electron density scale height to the radius of the neutron star, i.e., $(H_{\rm e}/R)^{-1}$.
As discussed above, the magnetic field grows until $\ell \to 1$, i.e., until the wavenumber approaches a dipole from the initial small-scale wavenumber $\ell_{\rm max}$.
For later reference, values of $\ell_{\rm max}=40$, 20, and 10 correspond to $\tau_2/\tau_1\approx10^4$, 1800, and 300.
Owing to the linear scaling of $B_{\rm LS}$ in \Eq{eq:BLS1}, the ratios $\tau_2/\tau_1$ also correspond to the amplification factor of the large-scale field.

\begin{figure*}
\centering
  \begin{tabular}{cc}     
        \includegraphics[width=0.95\textwidth]{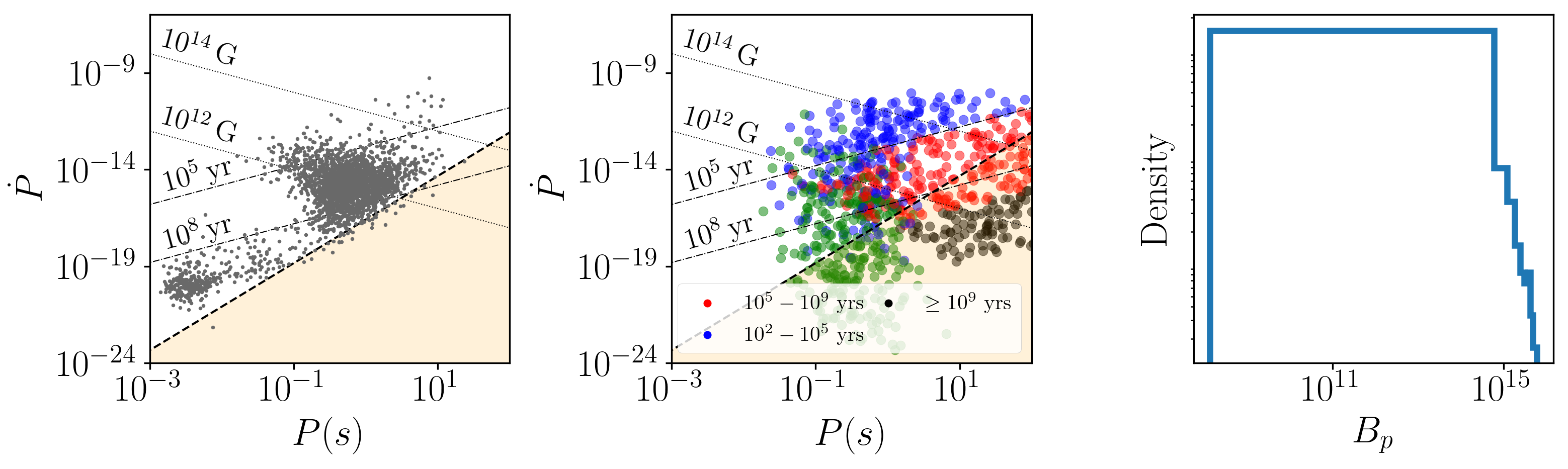} \hspace{-.4cm}
  \end{tabular}
  \caption{Real $P\dot{P}$ diagram of pulsars obtained from the ATNF catalog while the middle panel shows a simulated $P\dot{P}$ diagram with our model for a $1000$ different neutron star initial conditions using $\tau_1=10^2\yr$ and $\tau_2=10^6\yr$, corresponding to $\ell_{\rm max}=40$. Both panels include lines for constant magnetic fields and characteristic ages and the pulsar death line, calculated using the \texttt{psrqpy} software package. The color of the dots in the middle panel indicates the age of the selected neutron star, with neutron stars older than $\unit[10^9]{yr}$ in black, red for neutron stars between $\unit[10^5$--$10^9]{yr}$,
blue for neutron stars between $\unit[10^2$--$10^5]{yr}$, and green for neutron stars younger than $\unit[10^2]{yr}$.
The last panel shows the histogram of the large-scale magnetic field of the $1000$ neutron stars at their selected ages.}
\label{fig2}  
\end{figure*}
\section{Simulation}\label{sec:simulation}
With the model for the spin and magnetic field evolution of neutron stars described above, we now consider the effect of such a model on several neutron star properties and observables through a series of simulations.
For each of our simulations, we draw the initial spin-period $P_0$
and initial magnetic field $B_{\rm p,0}$ from astrophysically motivated distributions $p$ \citep{Igoshev2022, Pagliaro2023}, in particular,

\begin{align*}
p(\ln P_{0}/\unit{s}) &= \mathcal{N}(-1.25, 0.99) , \\
p(\log_{10} B_{\rm p,0}/\unit{G}) &= \mathcal{U}(8, 11),
\end{align*}
where $\mathcal{N}$ and $\mathcal{U}$ denote normal and uniform
distributions in given ranges.

We also fix $m = 2.3$ throughout this work, consistent with the decay of the magnetic field under ohmic dissipation~\citeg{Pons2007}.
We note here that a precise value of $m$ is not important to this work as we focus on the growth of the magnetic field rather than the decay.
We have verified that the results are robust to our choices of $m$, the initial spin, and large-scale magnetic field.
Our motivation for choosing these ranges were to show 1) that one does not need large scale magnetic fields at birth to make magnetars and 2) that a regular neutron star spin population can make magnetars.

We start by investigating the evolution with time; we set
$\tau_{1} = \unit[10^{2}]{yr}$ to $\tau_{2}=\unit[10^6]{yr}$, i.e., $\ell_{\rm max} \approx40$, and evolve $100$ neutron stars drawn with initial periods and magnetic fields from the distributions above forward in time.
Figure~\ref{fig:timeevolution} shows the evolution for different observables. 
In particular, we show the evolution of neutron stars across the $P\dot{P}$ diagram, the period and period derivative as functions of time, the observable braking index, dipole component of the magnetic field and the characteristic age.
Figure~\ref{fig:timeevolution} helps illustrate the model's main features.
For example, large-scale initial fields as small as $\unit[10^{11}]{G}$ can be amplified to magnetar strengths ($\unit[10^{14}]{G}$, alleviating the concern of needing a large-scale poloidal field at birth. In particular, the age at such field strengths depends on the choice of $\tau_{1}$ and $\ell_{\rm max}$. 
Similarly, the model can naturally produce a range of braking indices despite being built only on the assumption of vacuum-dipole emission, and naturally returns values of the braking index $n\lesssim 2$ for young pulsars, as measured in the observed population~\citep{2017Espinoza}.
We also show the evolution of the characteristic age with time. We find that at young ages, there is a huge discrepancy between the characteristic age and the actual age of the neutron star, while after some time these ages tend to agree. This is expected as our spin evolution is only different from  $n=3$ at early times.
The evolution across the $P\dot{P}$ diagram also demonstrates that the
model can accommodate neutron stars in various parts of the $P\dot{P}$
diagram despite a narrow starting position and explain the different
locations we observe for real neutron stars.

\begin{figure*}
\centering
  \begin{tabular}{cc}     
        \includegraphics[width=0.95\textwidth]{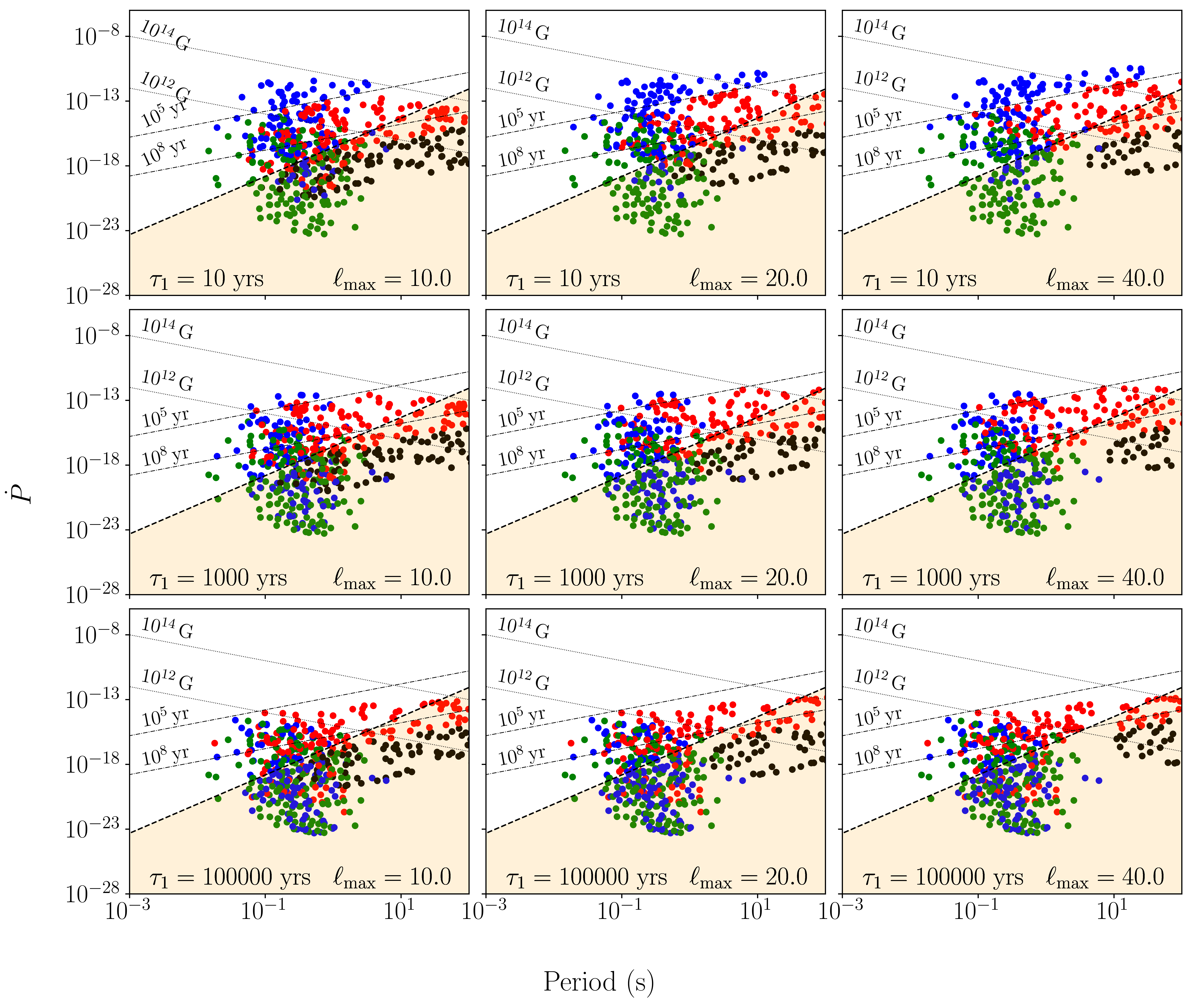} \hspace{-.4cm}
  \end{tabular}
  \caption{Simulated $P\dot{P}$ diagrams for the same initial conditions in all panels and ages but different $\tau_{1}$ and $\ell_{\rm max}$. The colors correspond to the ages of the neutron stars, with the same ranges as described in Fig.~\ref{fig2}.} 
\label{fig:ppdots}  
\end{figure*}

To illustrate this last point, we now simulate a hypothetical $P\dot{P}$
diagram with $\tau_{1} = \unit[1]{yr}$ and $\ell_{\rm max} = 40$
corresponding to $\tau_{2}=\unit[10^{4}]{yr}$, ignoring
selection effects and drawing ages randomly throughout the lifespan of the neutron star.
In Figure~\ref{fig2}, the left panel shows a real $P\dot{P}$ diagram of pulsars obtained from the ATNF catalog~\citep{Manchester2005} through the \texttt{psrqpy} software package~\citep{psrqpy}. 
The middle panel shows a $1000$ simulated neutron stars with our model drawn randomly from the above distributions and selected at different ages. 
Both panels include lines for constant magnetic fields and characteristic ages and the pulsar death line~\citep{Zhang2000}, calculated through \texttt{psrqpy} with the region below shaded in yellow. 
The last panel shows the histogram of the large-scale magnetic field of the $1000$ simulated neutron stars at their selected ages. 
We emphasize that we do not model processes such as mass transfer that is likely responsible for neutron stars observed in the lower left of the $P\dot{P}$ diagram, i.e., the millisecond pulsars. We caution the reader against a direct comparison of the observed population and our simulated population. We do not model the radio luminosity of any of the neutron stars and therefore cannot account for any selection effects that may limit our ability to see neutron stars in certain parts of the $P\dot{P}$ diagram~\citep{Faucher+06, Szary+14}.

Figure~\ref{fig2} also demonstrates that this evolutionary model for the magnetic field can produce neutron stars in the ``magnetar'' region of the $P\dot{P}$ diagram while also accounting for the distribution of neutron stars across the $P\dot{P}$ diagram for a small range of initial conditions.
Furthermore, it helps illustrate the significant differences in neutron stars' actual ages and magnetic fields compared to the characteristic spin-down age and constant magnetic-field models. 
This may be important to reconcile the often significant discrepancies in inferred ages of neutron stars in supernova remnants with their location in the $P\dot{P}$ diagram.

The above $P\dot{P}$ diagram represented one single case of $\tau_{1}$ and $\ell_{\rm max}$. However, as we mentioned in Section~\ref{sec:b_increase}, these values are unknown.
In Figure~\ref{fig:ppdots}, we show simulated $P\dot{P}$ diagrams for a range of these values.
We note that each panel has the same initial conditions (i.e., $P_0$ and $B_{\rm p,0}$) and all neutron stars are selected at the same ages, and the difference in their ``observed'' locations is only due to different $\tau_{1}$ and $\ell_{\rm max}$.
Figure~\ref{fig:ppdots} can be interpreted in two primary ways.
First, for almost any choice of $\tau_{1}$ and $\ell_{\rm max}$, this model can accommodate a large portion of the $P\dot{P}$ diagram for a small range of initial conditions and provide a natural explanation for the often discrepant age estimates from spindown and a supernova remnant. Second, to explain magnetars with this model, we need large $\ell_{\rm max}$ and small $\tau_{1}$, i.e., we need the inverse cascades to begin early and need the initial field to be confined to small scales. 

\section{Conclusions}\label{sec:conclusions}
Our results demonstrate that it is indeed possible to explain the
production of large-scale magnetic fields through an inverse cascade.
This means that an initial small-scale magnetic field gets gradually
converted into a large-scale one.
The efficiency of this process increases with increasing dynamic range
in space and time, i.e., it becomes more efficient, the smaller the
scale of the initial field (larger $\ell_{\rm max}$), and the earlier
the inverse cascade begins (smaller $\tau_1$).

Our models for the growth of the magnetic field early on in the life of the star can explain the low ($n\lesssim 2$) braking indices measured in young pulsars, and their observed evolution in the $P\dot{P}$ diagram~\citep{2017Espinoza}.
This is consistent with numerical results presented in \citet{GC15}, where the low braking indices of Vela and other young pulsars were interpreted as a byproduct of the increase in large-scale magnetic fields in Hall MHD. However, note that~\citet{GC15}, started with initially strong large-scale magnetic fields, while we can reproduce magnetars with much smaller initial large-scale fields. 

The most extreme case studied in this Letter is $\ell_{\rm max}=40$ and $\tau_1=1\yr$.
We estimated $\ell_{\rm max}=20$ based on the scale height, but this provided a rough estimate,
because the convective eddies during the first minute of the neutron star's life can well be smaller than the local scale height as discussed in Section~\ref{sec:b_increase}.
Somewhat larger values of $\ell_{\rm max}=40$ appear therefore feasible.
However, the situation with the value of $\tau_1$ is less obvious.
Once the crust has been established, the relevant timescale is
the magnetic diffusion time, which was also used as the natural time
unit in B20 and \cite{Bra23}.
In their Cartesian models, it was defined as $\tau_{\rm d}=\xiM^2/\eta$.
Using $\eta=4\times10^{-8}\m^2\s^{-1}$ and $\xiM=R/\ell_{\rm max}\approx700\m$ with $\ell_{\rm max}=20$, we have $\tau_{\rm d}=\unit[0.4]{Myr}$.
Somewhat smaller values are expected as we increase $\ell$, for example to 100,
in which case we can have $\xiM=150\m$, and therefore $\tau_{\rm d}=\unit[20]{kyr}$.
In addition, the time $\tau_1$ when the inverse cascade begins is a certain
fraction of $\tau_{\rm d}$, for example $\tau_1/\tau_{\rm d}=0.01$, corresponds to the time when
a certain characteristic quantity (the Hosking integral) reached a maximum;
see Figure~4 of \cite{Bra23}.
Thus, the smallest conceivable value of $\tau_1$ is 200~yr for a standard crustal composition.
It should be noted, however, that recent calculations of the conductivity in the presence of a nuclear pasta phase \citep{Pasta2023} have yielded significantly smaller conductivities, which could accommodate values of $\tau_1\approx 1$ yr. 

Our results show that low values of $\tau_1$ (together with high values of $\ell_{\rm max}$) will create stars in the magnetar range in most of our models, and therefore point towards
the presence of a thin, highly resistive pasta layer at the base of the crust, as suggested also by \cite{2013NatPh...9..431P},
to explain the absence of long period isolated X-ray pulsars.
We note that all models produce neutron stars below the death line, but those would not be observable as pulsars
owing to their low luminosity.

Another implication of this model is on the ages of magnetars. If the magnetic field is confined to small-scales at birth and the initial dipole component is small, then there needs to be a minimum time such that the field can grow to magnetar strengths.  For the choice of $\tau_1 =\unit[10]{yr}$ and $\ell_{\rm max}=40$, this is $\geq \unit[500]{yr}$ for our simulation, i.e., there should not exist any neutron stars with $B_{\rm p} \gtrsim \unit[10^{14}]{G}$, younger than $\approx \unit[500]{yr}$. A comparison between characteristic age and true age is presented in Figure~\ref{fig:timeevolution}, where we see that at early times most of our systems are much younger than they appear (we caution, however, that the systems we produce with large characteristic ages have very small values of $\dot{P}$ at birth, which would make them challenging to observe), while in the range between $10^3$ and $10^5$ years, a small number of systems, which correspond to the magnetars, are slightly older than they appear.   However, we note that we do not capture the true diversity of initial conditions such as spin or considerations of the equation of state which could produce magnetar-like field strength earlier in time. 
We will explore the full effect of different initial conditions, equations of state, and compare directly to observations in future work. 
\section{Acknowledgments}
We are grateful to Clara Dehman, Cristobal Espinoza, and Kostas Gourgouliatos for helpful comments.
We also thank the referee for their helpful feedback that has improved the presentation of this work.
N.S.\ is supported by a Nordita fellowship. 
A.B.\ acknowledges support from the Swedish Research Council (Vetenskapsr{\aa}det, grant number 2019-04234).
Nordita is supported in part by NordForsk. B.H. acknowledges support from the National Science Centre Poland (NCN) via grant OPUS 2018/29/B/ST9/02013.
\bibliographystyle{aasjournal} 
\bibliography{ref}
\end{document}